\documentclass{fullpaper_hutech}

\begin{document}

\title{Bifurcation and Nonlinear Oscillation of the Bead Motion}
\twocolumn[
\begin{@twocolumnfalse}
\begin{flushleft}
\maketitle
\begin{abstract}
\textbf{Sanghwa Lee}\\
\vspace{0.5cm}
torytony24@gmail.com\\
\vspace{0.5cm}
Korea Science Academy of KAIST, 20-078\\
\vspace{0.5cm}
\textbf{ABSTRACT}\\
This research focuses on the interesting physical phenomenon of the bead-hoop system. The bifurcation can be observed investigating the equilibrium point of the bead, and nonlinear oscillation also occurs from the bead's motion. This paper includes the theoretical investigation by setting the quantitative model and studying pitchfork bifurcation. Also, the equation of motion is derived by investigating resistance. Three parts of experiments were done and went through analysis by examining standard error, error of experiment, and the error of fitting.

\end{abstract}
\vspace{20pt}
\end{flushleft}
\end{@twocolumnfalse}
]
\thispagestyle{firstpage}

\lstset{
basicstyle=\ttfamily,
}

\section{INTRODUCTION}

The bead-hoop system is a simple system in a circular hoop that rotates about its diameter and a small spherical bead rolled in a groove on the hoop. Such a simple system shows interesting dynamics, especially the motion of the spherical bead. The investigation of the whole system will be focused on the motion of the bead in two parts, on the topic of equilibrium and the topic of oscillation. Analysis of each topic shows interesting, non-linguistic results. This paper shows the investigation of the system through quantitative analysis and verifies it through experiments and simulations.

\section{THEORY}

The investigation on the topic of equilibrium was firstly investigated as the hoop rotates about its diameter, the energy of the bead changes. This results from the bead having a new equilibrium point that has angular displacement about its initial equilibrium point. There could be one more degree of freedom that the axis of rotation having an angle with respect to the $xy$ plane. Considering all these factors, it is able to set up the quantitative model.

\subsection{Quantitative Model}

The bead-hoop system can be modeled as \textbf{Fig. 1}. The bottom end of the rotation axis of the hoop is located in origin. The axis has an angle with respect to the $z$ axis. To denote this angle, setting a unit vector $\hat{n}$ which has the same direction as the axis would help. For angle $\alpha$, it is denoted as
\begin{equation}
\begin{split}
\alpha=\arccos{(\hat{z}\cdot \hat{n})}
\end{split}
\end{equation}

Other parameters are considered, such as the angular displacement with respect to the initial equilibrium point; the lowest point of the hoop is denoted as $\theta$. Also, the radius of the hoop $R$, the radius of the bead $r$, the width of the groove $l$, depth of the groove $a$, the mass of the bead $m$. The angular velocity of the hoop will be shown as $\vec{\omega}=|\vec{\omega}|\hat{n}$, and angular speed as $|\vec{\omega}|=\omega$. Also, setting the angular displacement with respect to the coordinate system of the bead as $\phi$ is available.

\begin{figure}
\begin{center}
\includegraphics[width=7cm]{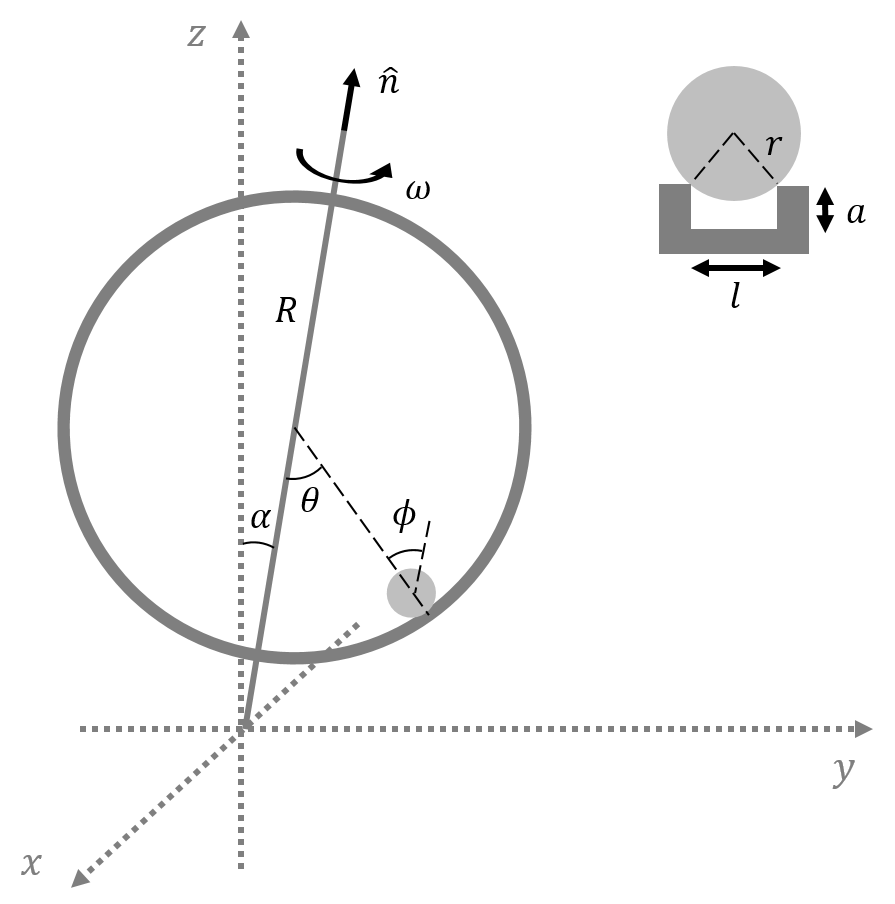}
\caption{Coordinate system}
\label{figure::label}
\end{center}
\end{figure}

\subsection{Investigation of Equilibrium}

As the hoop rotates about its axis, the equilibrium point of the bead moves. Considering the vertical case in which $\alpha=0$, it is able to draw a free body diagram as \textbf{Fig. 2}. By solving the equation of motion, we can get the stable equilibrium point as the following.

\begin{figure}
\begin{center}
\includegraphics[width=5cm]{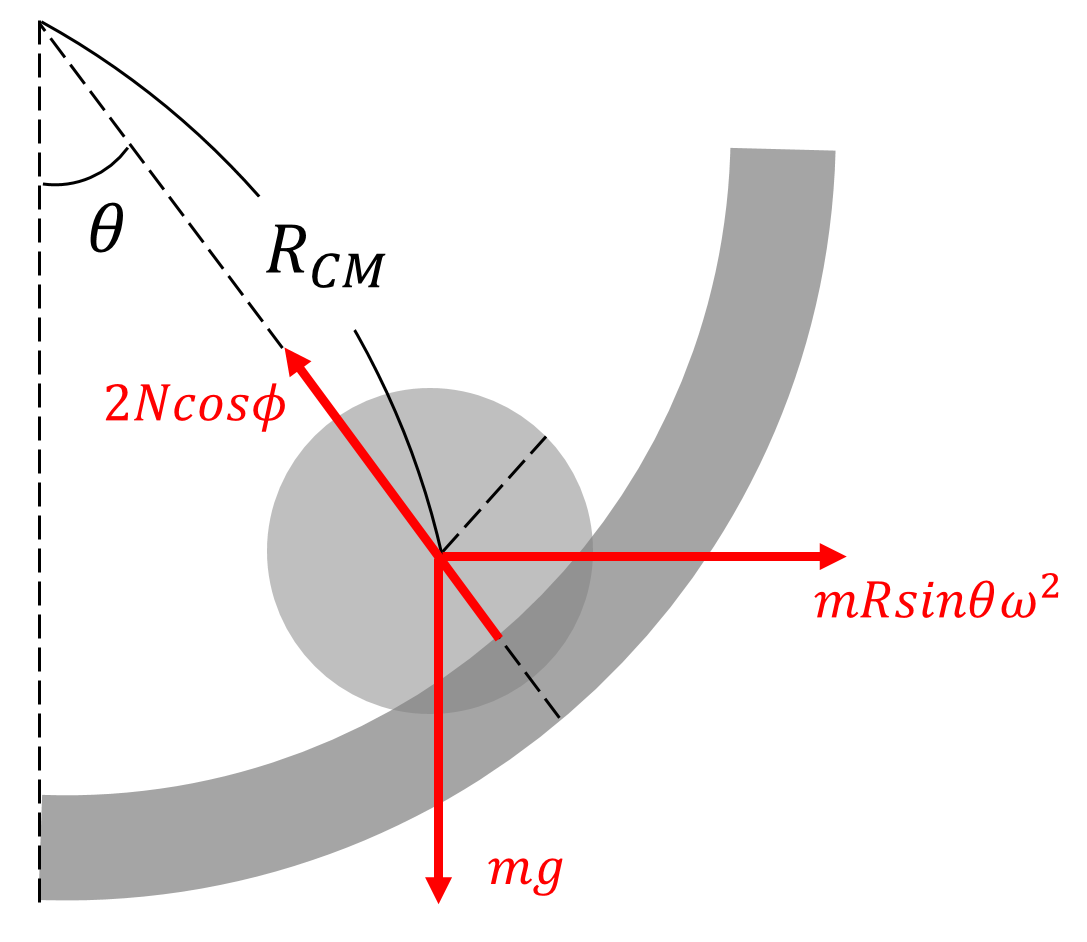}
\caption{Free Body Diagram}
\label{figure::label}
\end{center}
\end{figure}

\begin{equation}
\begin{split}
mg\sin{\theta}=mR_{CM} \sin{\theta} \cos{\theta} \omega^2\\
\sin{\theta}=0 \quad \textrm{or} \quad \cos{\theta}=\frac{g}{R_{CM} \omega^2}\\
\theta=0 \quad \textrm{or} \quad \theta=\arccos{\left( \frac{g}{R_{CM} \omega^2}\right)}
\end{split}
\end{equation}

The equilibrium point of the bead varies as the angular speed of the hoop changes. The criteria of this change can be investigated through the potential energy of the bead. Setting the critical angular speed $\omega_c$ as

\begin{equation}
\begin{split}
\omega_c = \sqrt{ \frac{g}{R_{CM}} }
\end{split}
\end{equation}

By integrating conservative force,

\begin{equation}
\begin{split}
U(\theta) = - \int \left[ -\frac{\omega_c^2}{\lambda} \cos{\theta} + \frac{\omega^2}{2\lambda} \cos^2{\theta} \right]
d \theta
\end{split}
\end{equation}

Graphing the potential energy $U(\theta)$ by $\theta$, it is able to see the equilibrium point. When observed through the graphing tool, the potential graph before the angular speed reaches $\omega_c$ shows a simple concave upward graph with one local minimum. However, the concave upward graph splits into two dimples so that two equilibrium exists. This has two local minimums and one local maximum for the unstable equilibrium.

Explanation with the bifurcation graph gives the same result. The bifurcation of the stable equilibrium point occurs as the angular speed $\omega$ goes over the critical angular speed $\omega_c$. This phenomenon is known as pitchfork bifurcation. Until the angular speed of the hoop reaches the critical angular speed $\omega_c$, the equilibrium point still stays on the initial equilibrium point. However, when the angular speed of the hoop reaches $\omega_c$, the bifurcation occurs. The equilibrium point moves along the hoop. Since the hoop is circular, the negative angle and positive angle can both be the new equilibrium point due to symmetry. This results in the fork-like graph when the angular displacement is graphed through the variation of the angular speed shown as \textbf{Fig. 3}.

\begin{figure}
\begin{center}
\includegraphics[width=6cm]{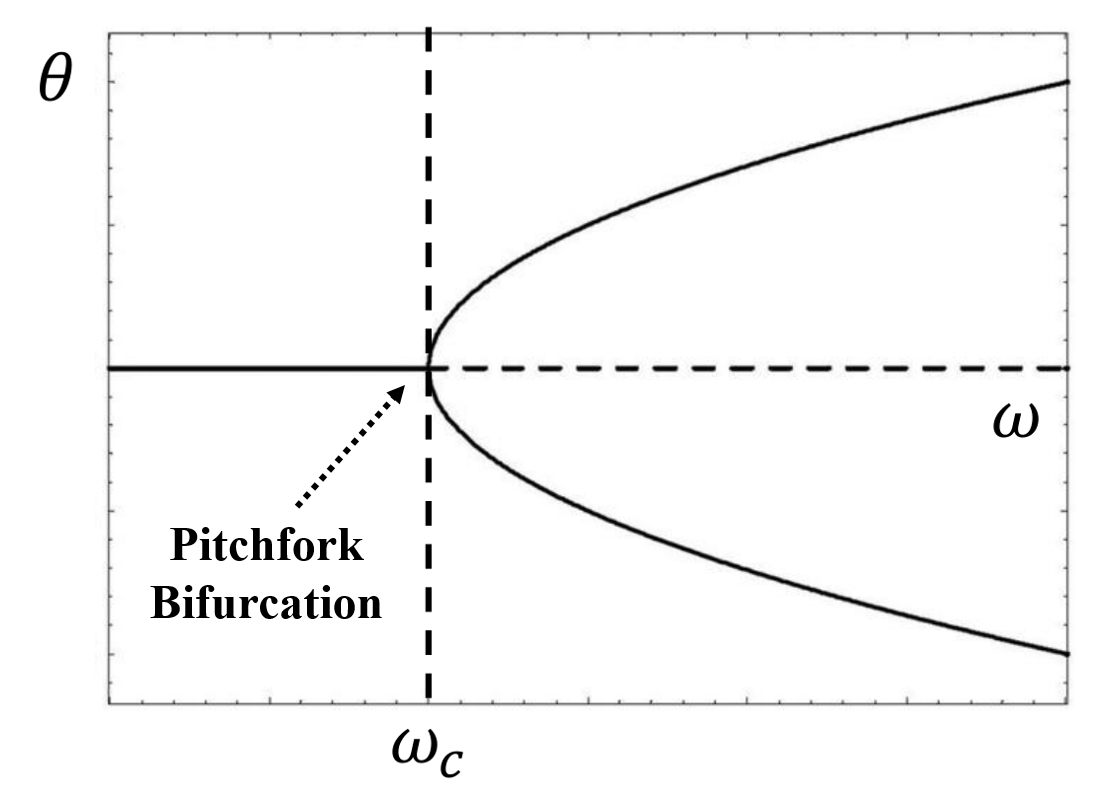}
\caption{Pitchfork Bifurcation}
\label{figure::label}
\end{center}
\end{figure}

\subsection{Investigation of the Oscillation}

In order to set the equation of motion of the bead, the resistance of the bead must be considered. The previous research \cite{main} have just assumed linear resistance. This research will have a more detailed explanation of resistance. The resistance that acts on the rolling bead can be divided into two types, air resistance, and rolling resistance. First, air resistance exists due to drag force. Since the bead is relatively small and spherical, it is able to apply Stokes' law. By Stokes' law, considering $\mu$ as the viscosity of air,

\begin{equation}
\begin{split}
F_d = 6 \pi \mu r v
\end{split}
\end{equation}

Where $v$ is the speed of the center of the bead, since $\mu$ and $r$ is constant, we can conclude about the air resistance that the drag force is proportional to the speed of the bead.

Also, in terms of rolling resistance, rolling friction is a force that makes resistance. The magnitude of the force is proportional to the normal force $N$ with a rolling constant $C_r$. The rolling constant is affected by three main factors, the roughness of the surface, radius $r$, and speed $v$. Two parameters, roughness and $r$, are constant. This results that the rolling friction is also proportional to velocity \cite{resistance}.

The total resistance is the addition of drag force and rolling friction, which the following equation satisfies for resistance constant $b$.

\begin{equation}
\begin{split}
F_{res} = F_d + F_r = bv
\end{split}
\end{equation}

Now it is able to set up the equation of motion. By using the parameter that was defined, the following equation satisfies by the rolling boundary condition.

\begin{equation}
\begin{split}
\dot{\phi} = \frac{R_{CM}}{r} \dot{\theta}
\end{split}
\end{equation}

In order to use the Euler-Lagrangian equation, kinetic energy $T$ can be calculated as the following.

\begin{equation}
\begin{split}
T&=\frac{1}{2} m ( R^2_{CM} \dot{\theta} + R^2_{CM} \omega^2 \sin^2{\theta} ) + \frac{1}{2} I \dot{\phi}\\
&=\frac{1}{2}mR^2_{CM} ( \dot{\theta}^2 +\omega^2 \sin^2{\theta}) + \frac{1}{2} \frac{2}{5} m \frac{r^2 (R-a)^2}{r^2 - \left( \frac{l}{2} \right)^2} \dot{\theta}^2\\
&=\frac{1}{2} m R^2_{CM} \left[ \left( 1+ \frac{2}{5} \frac{ r^2 (R-a)^2 }{R^2_{CM} \left( r^2-\frac{l^2}{4} \right) } \right) \dot{\theta}^2 \omega^2 \sin^2{\theta} \right]
\end{split}
\end{equation}

Setting the geometric constant $\lambda$ as

\begin{equation}
\begin{split}
\lambda= 1+ \frac{2}{5} \frac{ r^2 (R-a)^2 }{R^2_{CM} \left( r^2-\frac{l^2}{4} \right) }
\end{split}
\end{equation}

Then also calculating potential energy $V$, it is able to calculate Lagrangian $L$.

\begin{equation}
\begin{split}
T&= \frac{1}{2} mR^2_{CM} ( \lambda \dot{\theta}^2 + \omega^2 \sin^2{\theta})\\
V&=-mgR_{CM}\cos^2{\theta}\\
L&=\frac{1}{2}mR^2_{CM} ( \lambda \dot{\theta}^2 + \omega^2 \sin^2{\theta}) +mgR_{CM} \cos{\theta}
\end{split}
\end{equation}

Using the Euler-Lagrangian equation, it is able to get the equation of motion.

\begin{equation}
\begin{split}
\frac{\partial L}{\partial \theta} &- \frac{d}{dt} \left( \frac{\partial L}{\partial \dot{\theta}} \right) =0\\
\ddot{\theta} + \frac{g}{R_{CM}\lambda} \sin{\theta} &-\frac{R_{CM} \omega^2 }{\lambda} \sin{\theta} \cos{\theta} =0
\end{split}
\end{equation}

Resistance can also be considered by adding the term. Also, since the rotational axis has angle $\alpha$ with respect to the $z$ axis, the driving force $F(\theta)$ exists on the left-hand side of the equation.

\begin{equation}
\begin{split}
\ddot{\theta} + \frac{b}{\lambda m} \dot{\theta} + \frac{1}{\lambda} \left( \frac{g}{R_{CM}} -\omega^2 \cos{\theta} \right) \sin{\theta} = F(\theta)
\end{split}
\end{equation}

The gravitational force makes the drive to the bead. Using geometry, we can get the final equation of motion,

\begin{equation}
\begin{split}
\ddot{\theta} &+ \frac{b}{\lambda m} \dot{\theta} + \frac{1}{\lambda} \left( \frac{g}{R_{CM}} \cos{\alpha} -\omega^2 \cos{\theta} \right) \sin{\theta} \\&= \frac{1}{\lambda} \frac{g}{R_{CM}} \sin{\alpha} \cos{\omega t} \cos{\theta}
\end{split}
\end{equation}

The final equation of motion shows the forced nonlinear oscillation. The oscillation around the equilibrium point $\theta_0$ would lead the equation linearized, and by shifting the equation by $\theta$, we can get the following linearized equation.

\begin{equation}
\begin{split}
\ddot{\theta} + \frac{b}{\lambda m} \dot{\theta} + \frac{1}{\lambda} \left( \frac{g}{R_{CM}} \cos{\alpha} -\omega^2 \right) \theta = \frac{1}{\lambda} \frac{g}{R_{CM}} \sin{\alpha} \cos{\omega t}
\end{split}
\end{equation}

This equation is a general form of forced damped harmonic oscillator. The amplitude of this oscillator can easily be calculated as the following.

\begin{equation}
\begin{split}
A=\frac{\alpha g}{R_{CM} \sqrt{ \left( (1+\lambda)\omega^2 - \omega_c^2 \right)^2 + \frac{b^2}{m^2}\omega^2 }}
\end{split}
\end{equation}

\section{EXPERIMENT}

Two main subjects, pitchfork bifurcation and oscillation resonance are discovered. Experiments to verify these subjects will be done. Also, the inclined angle $\alpha$ can be calculated through the experiment. One of the assumptions is that the bead never rolls out of the groove nor gets lifted by inertial force. Two contact points with the bead and the groove will always be contacted throughout the whole experiment. For the bead to be contacted on both sides, the inclined angle $\alpha$ must be small enough. Small-angle $\alpha$ cannot be measured directly, so by analyzing the result of the experiment, $\alpha$ can be measured reversely.

\subsection{Experimental Setting}

\begin{figure}
\begin{center}
\includegraphics[width=8cm]{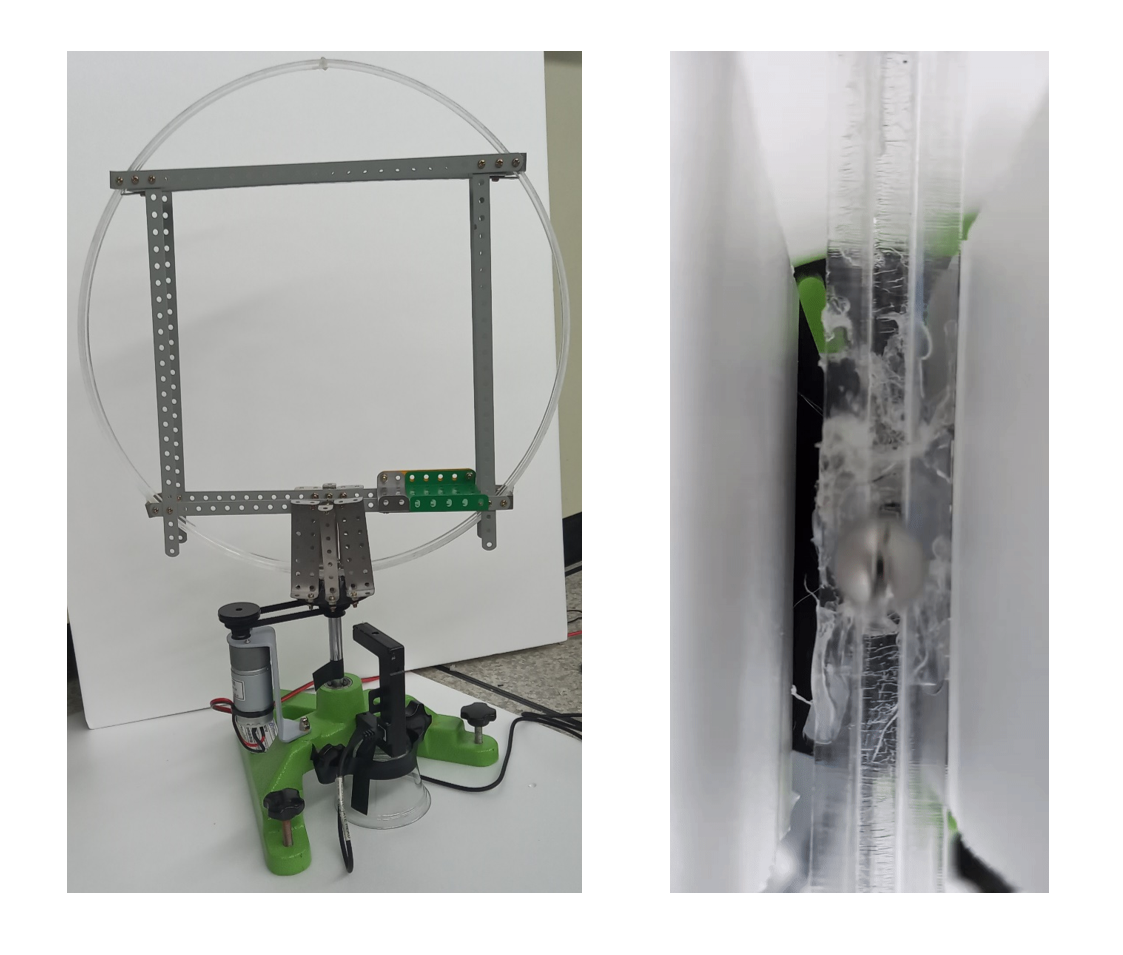}
\caption{Experimental apparatus / View from the camera}
\label{figure::label}
\end{center}
\end{figure}

As shown in \textbf{Fig. 4}, the experimental setting has a hoop with a motor connected by a belt. The motor and the hoop are firmly set on a stand, which the angle with respect to the $xy$ plane can differ. The voltage on the motor can be varied by PASCO Capstone Interface 850, and the angular speed of the hoop varies as the voltage changes. In order to measure the angular speed of the hoop, a photogate was set parallel to the $z$ axis. The hoop is made with an acrylic plate, and it is firmly set into the circle by cutting and holding it with metal bars. Also, the camera stand is set on the vertical metal bar so that it spins with the hoop. The camera moves along the inertial coordinate of the hoop so that the oscillation of the bead can be detected.

\subsection{Experiment of Equilibrium}

To measure the position of the equilibrium point, the video was taken at a distance of 1.5 m so that it is far enough to neglect the distortion from the camera. Since beads oscillate near the stable equilibrium, it is difficult to find the exact equilibrium position. The result was token ten times and was calculated the average. Starting with 3 rad/s, the equilibrium was measured for each speed by increasing by 0.2 rad/s.

In the experimental graph, the angular displacement of the bead from the initial equilibrium point varying the angular speed of the hoop is shown as \textbf{Fig. 5.} Experiments have clearly shown that theta becomes positive, and bifurcation occurs when the angular velocity of the ring exceeds the critical angular velocity. If the angular velocity is above the critical angular velocity, it follows the form of the inverse of cosine shown in the theoretical analysis. Also, the small error bar and the fact that the theoretical fitting fits in error bars clearly show that the experiment is done well. The reason why \textbf{Fig. 5.} doesn't show the negative theta part of the pitchfork bifurcation graph is that measured $\theta$ is only measured for the absolute value of the angle as $|\theta|$. By symmetry, the graph shows the exact pitchfork bifurcation of the bead-hoop system. The regression equation for this experimental data is shown in the following equation with fitting parameter $C_1$.

\begin{equation}
\begin{split}
\theta= \arccos{ \left( \frac{C_1}{\omega^2} \right)}
\end{split}
\end{equation}

\begin{figure}
\begin{center}
\includegraphics[width=8cm]{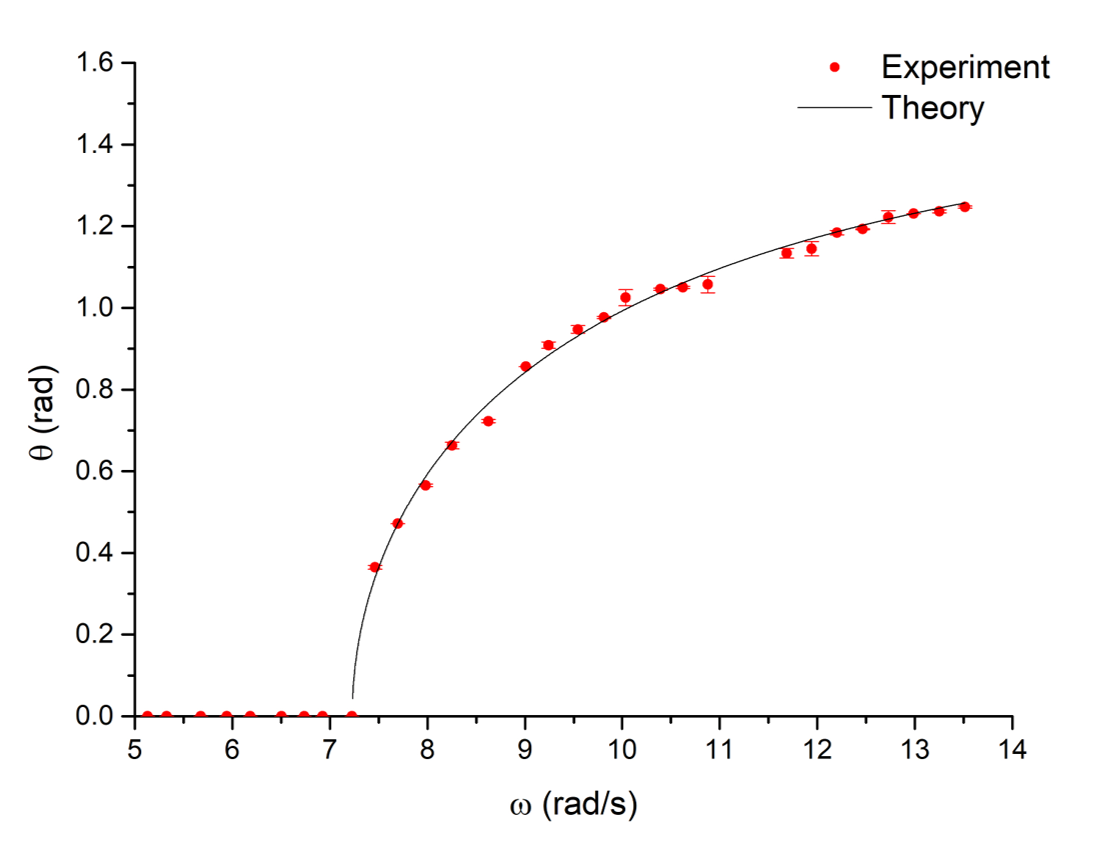}
\caption{$\omega$-$\theta$ Graph}
\label{figure::label}
\end{center}
\end{figure}

\subsection{Experiment of the Oscillation}

Before the nonlinear oscillation is investigated, the linear oscillation should be first investigated. Two main quantities that would be considered are amplitude and period. The equation that was derived for amplitude in the case of linear oscillation can be used.

In order to use the equation, the damping coefficient in the resistance term should be characterized by simply oscillating the bead when $\omega=0$ and $\alpha=0$ and plotting the data by time and amplitude, it is able to get the coefficient. The upper envelope and the lower envelope for the motion were calculated. It was able characterize the fact that $b/2m=0.187$.

Just as the experiment for the equilibrium, $\omega$ was varied, and the amplitude of the oscillation was measured. Since the linearized equation of motion shows the form of the forced damped harmonic oscillator, the resonance exists. As shown in \textbf{Fig. 6}, the experimental data were fitted using the following fitting equation with fitting parameter $C_2$.

\begin{figure}
\begin{center}
\includegraphics[width=8cm]{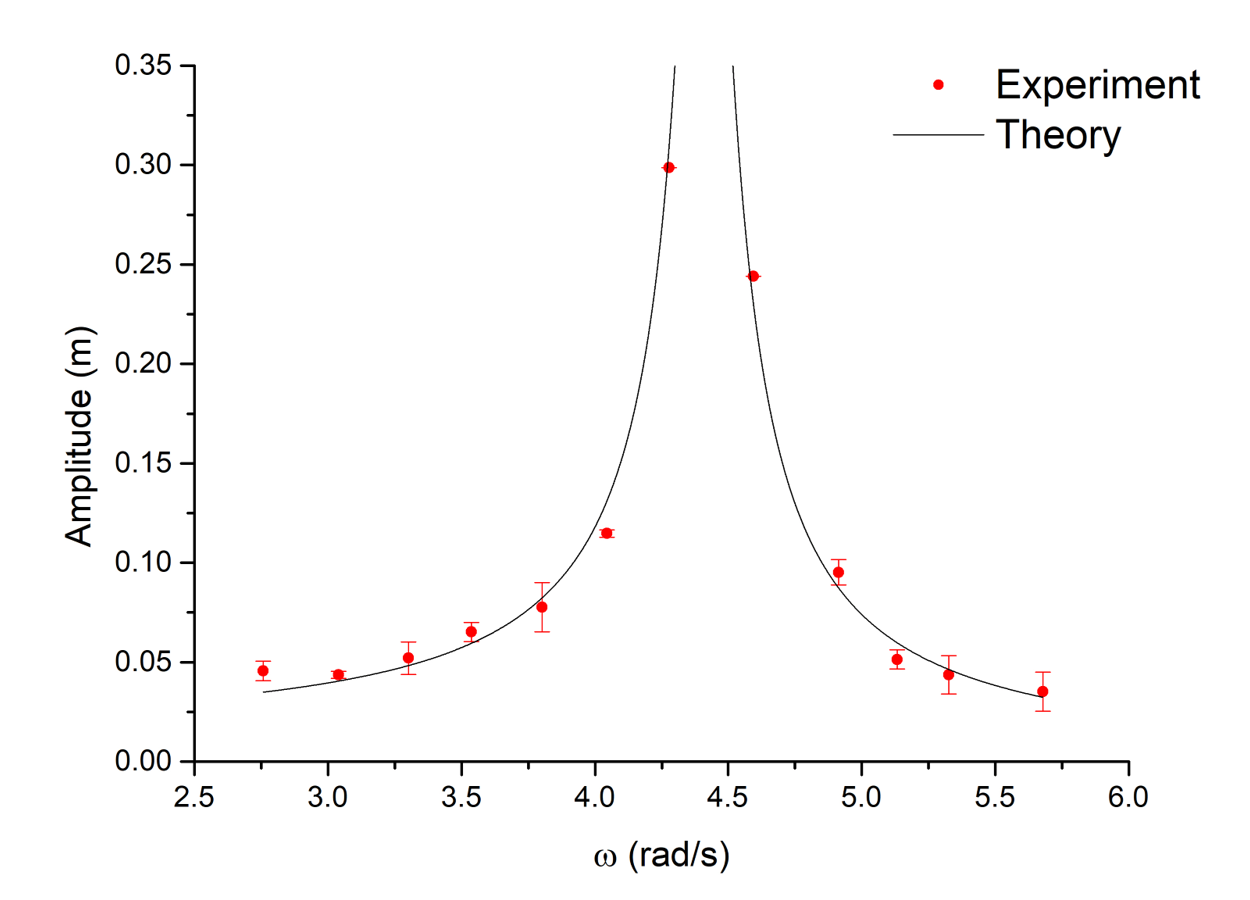}
\caption{$\omega$-$A$ Graph}
\label{figure::label}
\end{center}
\end{figure}

\begin{equation}
\begin{split}
A=\frac{C_2}{R_{CM} \sqrt{ \left( (1+\lambda)\omega^2 - \omega_c^2 \right)^2 + \frac{b^2}{m^2}\omega^2 }}
\end{split}
\end{equation}

Also, the oscillation period $T$ was measured and was compared with the theory. Just as the experiment for amplitude, the angular speed of the hoop increased, and $T$ was measured. As expected, the oscillating period got lower when $\omega$ increased. This experimental data has an inversely proportional regression equation as $T=2\pi / \omega$, so the data will be fitted with the following equation with $C_3$ as a fitting parameter.

\begin{equation}
\begin{split}
T=\frac{C_3}{\omega}
\end{split}
\end{equation}

Both graphs of amplitude and period as the vertical axis are shown in \textbf{Fig. 6} and \textbf{Fig. 7}. They show great coincidence between the experiment and theory. Also, the error bars are small, which means the experiments were done with high coherence.

\begin{figure}
\begin{center}
\includegraphics[width=8cm]{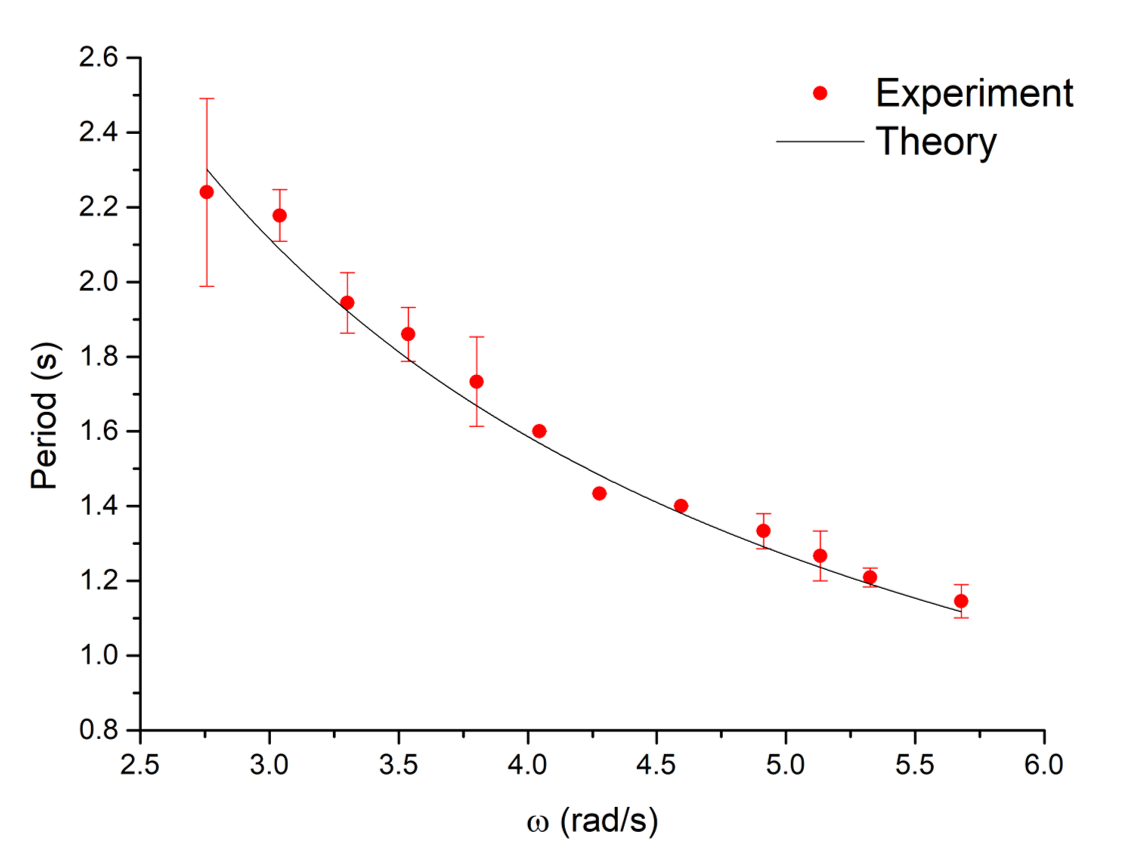}
\caption{$\omega$-$T$ Graph}
\label{figure::label}
\end{center}
\end{figure}

\section{ANALYSIS}

\subsection{Analysis of Equilibrium}

In the section of the experiment, the experiment of the equilibrium was done at \textbf{3.2}. As shown in \textbf{Fig. 5}, the experiment and the theory show great coincidence. The accuracy of this experiment can be analyzed numerically. Standard error $\sigma$ for this experiment results $\sigma=0.0469$, which is relatively small. Also, using equation (16), it is able to result that the fitting parameter $C_1=53.29$ neglecting the units. Theoretically, $\omega_c^2$ fits in the place of $C_1$. Plugging in the measurements such as the mass of the bead $m$, theoretical comparison parameter results $\omega_c^2=49.03$. By comparing two comparison values, $C_1$ and $\omega_c^2$, it is able to get the result that the error of the experiment is $7.99\%$. Also, by the following equation, the error of fitting can be calculated.

\begin{equation}
\begin{split}
\frac{\sigma}{C_1}\times 100(\%) = 0.09(\%)
\end{split}
\end{equation}

All the errors, standard error $\sigma$, the error between the experimental data and the theory, and the error of fitting are small enough to show that the experiment is done well-fitting the theory.

\subsection{Analysis of the Oscillation}

The experiment of the oscillation was done at \textbf{3.3}. For the amplitude experiment, the fitting was done using equation (17). Standard error $\sigma$ for this experiment is $\sigma=0.01284$, which is a small value relative to the magnitude of the data. Fitting parameter $C_2$ results as $C_2=0.99446$. It is able to calculate the error of fitting through the following equation.

\begin{equation}
\begin{split}
\frac{\sigma}{C_2}\times 100(\%) = 1.29(\%)
\end{split}
\end{equation}

The theoretical value of the fitting parameter $C_2$ is impossible to measure directly, so the purpose of this experiment is to reversely get the angle $\alpha$. Since the theoretical value that fits in the place of $C_2$ is $\alpha g$. $g$ value is well known, so finding out the value of $\alpha$ results in $\alpha=1.42^\circ$.

The experiment for the oscillating period also fits well with the theory. Standard error $\sigma$ for this experiment was 0.04641, and $C_3$ results $C_3=3.6457$. This value is $2\pi$ as theory. Calculating the error of the experiment, it is 0.98\% which is under 1\%. The error of fitting is also under 1\%, as shown in the following.

\begin{equation}
\begin{split}
\frac{\sigma}{C_3}\times 100(\%) = 0.731(\%)
\end{split}
\end{equation}

\subsection{Nonlinear Oscillation}

Equation (13) shows the equation of motion. The nonlinear oscillation has interesting phenomenon such as jump and hysteresis. The resonance graph, $\omega$-$A$ graph leans to the side as the nonlinearity increases. The nonlinearity of the equation refers to $\alpha$. As $\alpha$ increases, the nonlinearity increases, and the peak of the resonance graph tilts to the left side. In order to find out how the nonlinearity affects the motion, numerical analysis was done.

\begin{figure}
\begin{center}
\includegraphics[width=6cm]{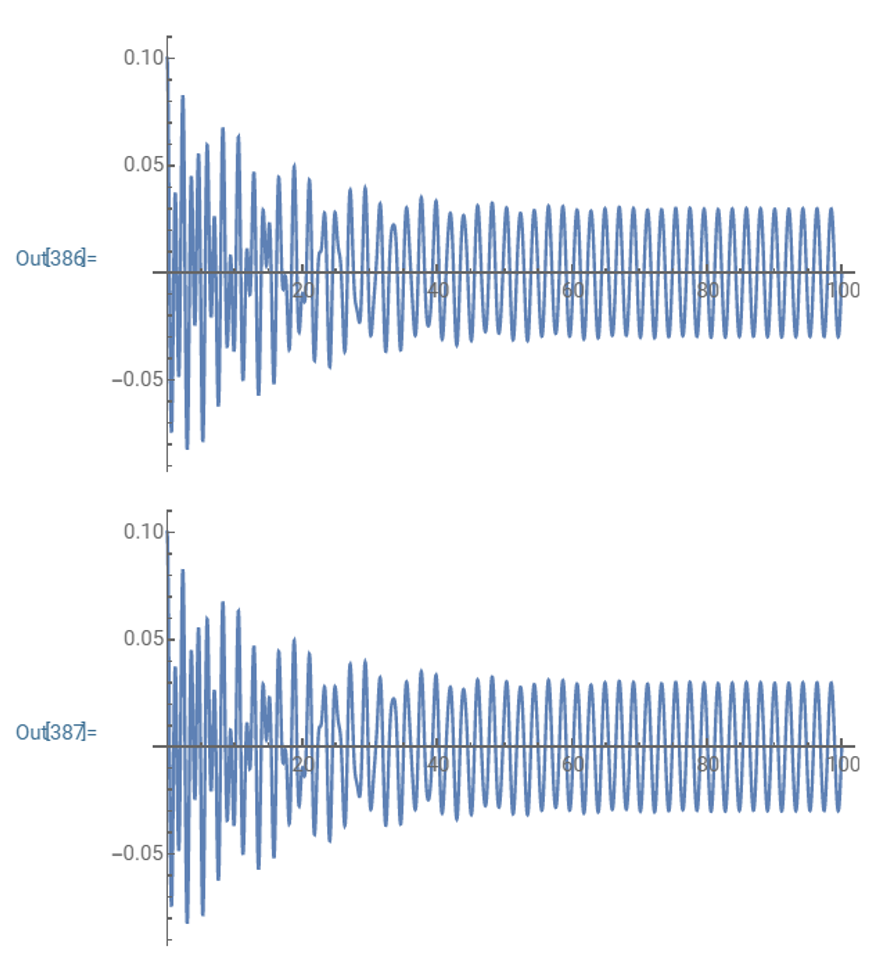}
\caption{Nonlinear / Linear motion}
\label{figure::label}
\end{center}
\end{figure}

As shown in \textbf{Fig. 8}, \textit{Out[386]} shows the nonlinear motion and \textit{Out[387]} shows the linear motion. The two motions don't show any significant visual difference. In order to compare two motions specifically, the nonlinear resonance graph can be drawn by Matlab. For the case when $\alpha=1^\circ$ as the experimental result, the graph hasn't shown any significant difference with the linear resonance $\omega$-$A$ graph. However, when $\alpha$ was over $7^\circ$, The hysteresis showed significantly as \textbf{Fig. 9}.

\begin{figure}
\begin{center}
\includegraphics[width=8cm]{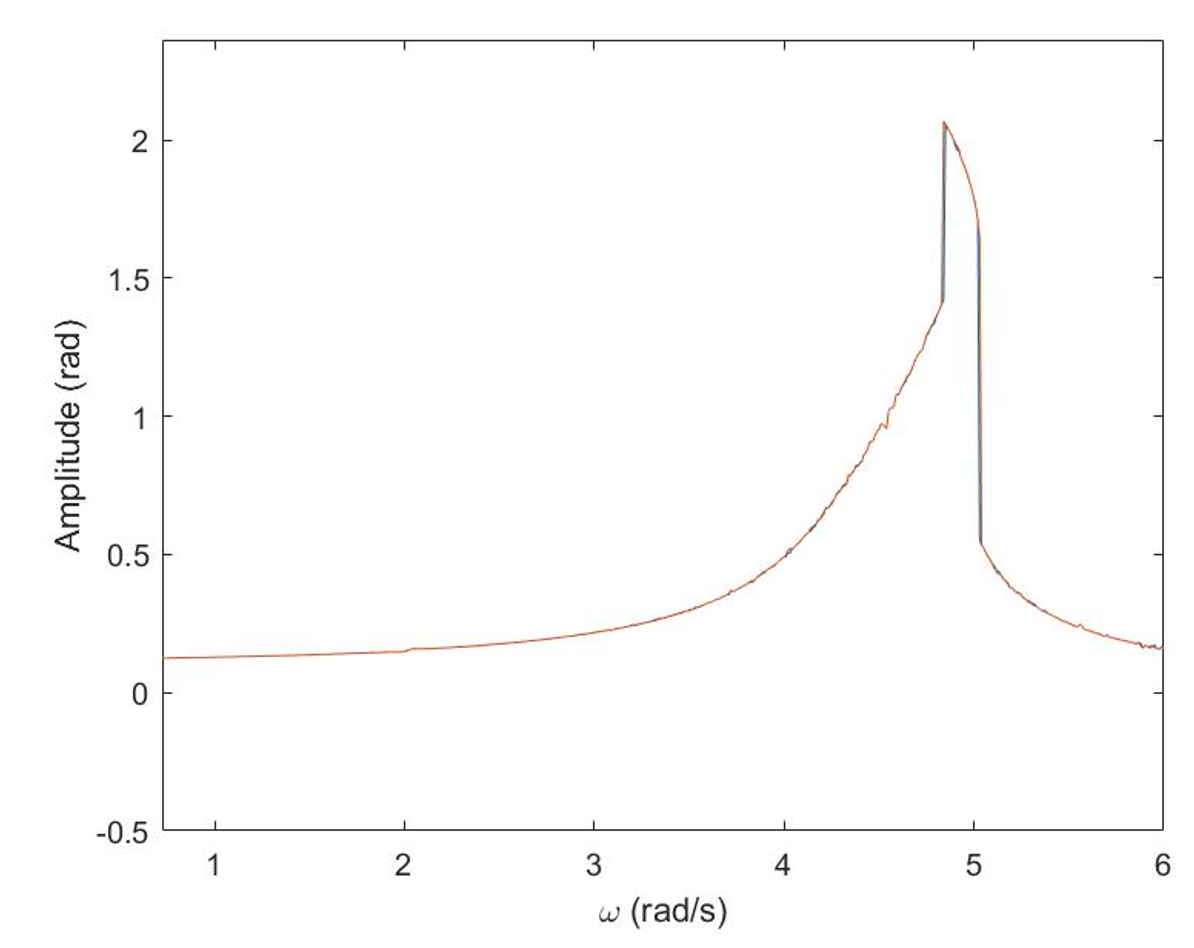}
\caption{Hysteresis when $\alpha=7^\circ$}
\label{figure::label}
\end{center}
\end{figure}

As shown in the figure, the resonance graph looks different when the nonlinearity is large. As the angular speed $\omega$ increases, the amplitude increases as usual. When $\omega$ reaches near the resonance angular frequency (angular speed), The amplitude increases at a sudden with discontinuity. Once it reaches the peak, the discontinuity occurs again, and amplitude decreases. This phenomenon shows the jump of the amplitude. When $\omega$ reversely goes from high value to low value, the amplitude value goes in a different pathway. It means that the value depends on the direction of the change. This phenomenon shows hysteresis. In this experiment, $\alpha$ was about $1^\circ$ so that jump and hysteresis cannot be significantly visualized.

\section{CONCLUSION}

This research went through theory, experiment, and analysis. The quantitative model was set as \textbf{Fig. 1}, and pitchfork bifurcation was explained by investigating the equilibrium and integrating conservative force. Also, finding out the fact that the resistance is proportional to velocity, the equation of motion was derived. The bead-hoop system was constructed based on the model, and an experiment of the equilibrium and oscillation was done. By analyzing the results, all the standard error, error of experiment, error of fitting was small enough to argue that the theory and experiment well match. The nonlinearity was also investigated and concluded that it is negligible for this experimental environment.

\end{document}